\documentclass[AMA,STIX1COL]{WileyNJD-v2} 
\articletype{}
\usepackage{moreverb}
\usepackage{amsmath}
\usepackage{amsfonts}
\usepackage{amssymb,xcolor}
\usepackage{soul}

\newcommand{\beq}{\begin{equation}}
\newcommand{\eeq}{\end{equation}}

\newcommand{\bra}[1]{\langle#1|}

\newcommand{\ket}[1]{|#1\rangle}

\newcommand{\id}{\leavevmode\hbox{\small1\normalsize\kern-.33em1}}

\newcommand\BibTeX{{\rmfamily B\kern-.05em \textsc{i\kern-.025em b}\kern-.08em
T\kern-.1667em\lower.7ex\hbox{E}\kern-.125emX}}

\begin{document}

\title{Improved Tomographic Estimates by Specialised Neural Networks}

\author[1]{Massimiliano Guarneri }

\author[2]{Ilaria Gianani}

\author[2,3]{Marco Barbieri}

\author[1]{Andrea Chiuri}

\authormark{GUARNERI \textsc{et al}}

\address[1]{\orgname{ENEA - Centro Ricerche Frascati}, \orgaddress{Via E. Fermi 45, 00044, Frascati, \country{Italy}}}

\address[2]{\orgdiv{Dipartimento di Scienze}, \orgname{Universit\`{a} degli Studi Roma Tre} , \orgaddress{Via della Vasca Navale 84, 00146, Rome, \country{Italy}}}

\address[3]{\orgdiv{Istituto Nazionale di Ottica}, \orgname{CNR}, \orgaddress{Largo E. Fermi 6, 50125, Florence \country{Italy}}}

\corres{Andrea Chiuri. \email{andrea.chiuri@enea.it}}

\abstract[Abstract]{Characterization of quantum objects, being them states, processes, or measurements, complemented by previous knowledge about them is a valuable approach, especially as it leads to routine procedures for real-life components. To this end, Machine Learning algorithms have demonstrated to successfully operate in presence of noise, especially for estimating specific physical parameters. Here we show that a neural network (NN) can improve the tomographic estimate of parameters by including a convolutional stage. We applied our technique to quantum process tomography for the characterization of several quantum channels. We demonstrate that a stable and reliable operation is achievable by training the network only with simulated data. The obtained results show the viability of this approach as an effective tool based on a completely new paradigm for the employment of NNs operating on classical data produced by quantum systems.}

\keywords{Quantum Channels, Neural Networks, Process Tomography} 

\maketitle

\section{Introduction}
Routine characterisation of quantum objects in a laboratory takes advantage from the fact that samples can often be collected in 
large numbers. This prevents fluctuations in the recorded counts to hamper the reconstruction: in a frequentist view, the experimental measurement frequencies can be made to match the probabilities within
the desired precision. Good level of control also ensures the object itself to belong to a known class, allowing to apply constraints and restricted methods for its characterisation~\cite{PhysRevA.66.062305,PhysRevLett.91.227901,PhysRevLett.92.087902,Branderhorst_2009,Blandino_2012,PhysRevA.84.050301}. 

For real-life quantum devices, however, these assets are hindered. A limited number of copies can be collected in reasonable times. Direct reconstruction methods are known to suffer from these fluctuations, and may even result in nonphysical outcomes: maximum likelihood algorithms help excluding these pathological instances~\cite{James01,Lvovsky_2004,Anis_2012}, but do not warrant an unbiased output with small samples. 

Oftentimes, the characterisation problem is cast in terms of effective parameters pertaining to the sought effect; for instance, we can be interested in assessing the depolarising action of a quantum channel, regardless of its exact form. In this way, specific strategies could be evaluated and put into action to curtail this specific detrimental effect. These parameters can indeed be retrieved by direct measurement - also when they refer to nonlinear properties - however, this technique does not benefit from the regularisation offered by full reconstruction. Thus, these are often obtained by a best-fit of the state or process matrices, when these are available.

The development of efficient and robust methods for estimating parameters in quantum processes requires efficient ways for dealing with finite measurement statistics and to avoid the uncertainty on the experimental frequencies. Machine learning (ML) techniques have been successfully adopted to tackle this challenge for black-box characterisation of quantum objects, with realisations including state and unitary reconstruction~\cite{spag17sr,gao18prl,torlai18nat,rocc19sa,torlai19prl,gior20prl,palmieri20npj,tiun20opt}, validation of quantum technology~\cite{agre19prx,flam19qst}, identification of quantum features~\cite{gebh20prr,cimini20prl}, and the calibration of quantum sensors ~\cite{hent10prl,cimi19prl,cimini21pra}. The versatility of ML has allowed to extend beyond these scopes to the adaptive control of quantum devices~\cite{hent11prl,love13prl,bona16nat,pali17neuro,liu17pra,paes17prl,lumi18pra,pali19pra,dina19prb,liu20mlst,peng20pra,ramb20prr}, as well as to the automated design of quantum experiments~\cite{knot16njp,arra19qst,nich19qst,meln18pnas,kren16prl,dris19qmi,saba19pra,gao20prl}. Realising this vision in full requires further exploration to understand the capabilities and opportunities offered by progress in ML~\cite{jaeg21nce,muja21}.

Considering our specific problem, the outcome of a tomography experiment is a collection of state, process, or detector matrices, obtained from the measured counts and their uncertainties; a common technique consists in simulating multiple events by Monte Carlo routines. Errors on the sought parameters are evaluated by best fit on each matrix of the Monte Carlo sample.
However, treating these matrices individually does not account for correlations among different elements of the matrix, while these could provide extra means for regularisation, a possibility currently under intense scrutiny~\cite{Schmale22}.

In this article, we discuss an extension of ML techniques to improve the accuracy in the tomographic estimation of parameters. A specialised neural network (NN) structure is built drawing inspiration from denoising procedures for imaging: indeed, each element of a process matrix can be thought as a pixel, thus incorporating convolutional layers in the NN can efficiently address their correlations. We illustrate these ideas with numerical examples of single- and two-qubit channels to suppress the impact of the statistics. The same techniques are also applied to real data, which also include experimental deviations. We demonstrate that it is possible to get useful results by training the NNs exclusively with a simulated dataset, showing generalization also to cases not seen during training.

\section{Methods}

In this article, we consider the action of a quantum channel, but a similar approach applies equally well to states and detectors. 
Data is collected and then processed, for instance by maximum likelihood methods, to deliver a matrix associated to the process. The uncertainties on the counts is accounted for by generating different outputs matrices based on either repeated measurements or, more often, on bootstrapping. This means that there is a collection of outputs available for further inspection.
\begin{figure*}[h]
\begin{center}
\includegraphics[width=\textwidth]{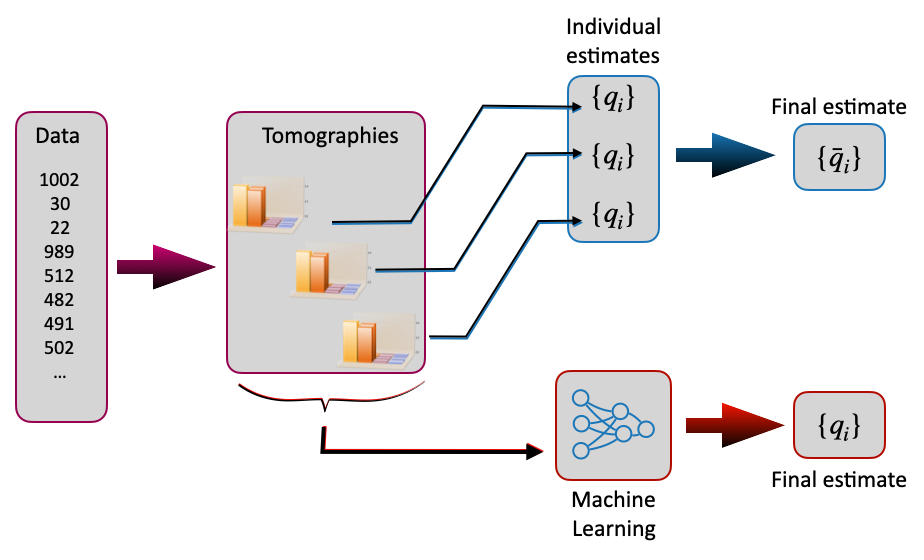}
\caption{General scheme of tomographic estimation of the parameters $\{q_i\}$. These are obtained by inspecting the output of matrix reconstruction generally by processing the individual instances one by one. We consider the performance of a Machine Learning procedure trained to account for the expected variability of the data.}
\label{fig_resume}
\end{center}
\end{figure*}

We then wish to isolate relevant features: this means we are interested in providing a description as a map $\Gamma_{\{q_i\}}$, where the set of parameters $q_i$ captures the essentials of our description. Values of these parameters by means of a tomography are what we term tomographic estimates. In practice, however, the experimental reconstruction may lead to a different map $\Gamma'$, due to the presence of unexpected effects or simply as a consequence of the finite statistics. Our aim is then to extract viable information on effective parameters $q_i$, {\it i.e.} on the physical effect they describe, from $\Gamma'$, while accounting for the variability of the data and attempting to curtail the consequent bias.

In the traditional approach, each output matrix delivers a different estimate for $\{q_i\}$, and these are then combined to obtain a final estimate as the average $\{\bar q_i\}$. In fact, a best fit procedure, minimising the distance between $\Gamma_{\{q_i\}}$ and $\Gamma'$, would be affected by the aforementioned discrepancies.

ML techniques may help tackling these issues. NNs can be trained to account for statistical fluctuations, and can learn to associate a matrix to its parameters within the expected variability. Since the uncertainty of any specific measurement has a nontrivial influence over all the elements of the process matrix, then it would be beneficial to include the presence of correlations explicitly in the network architecture.
This translates into introducing an autoencoder stage filtering the inputs. This also presents the advantage of limiting the detrimental effects of possible artefacts due to the maximum likelyhood estimation (MLE) reconstruction, and constitutes the main novelty of our approach.

\subsection{Quantum channels}
The action of a generic channel operating on a single qubit can be written as $\varepsilon [\rho] = \sum^3_{i,j=0}
\chi_{ij} \sigma_i \rho \sigma_j$ where the matrix $\chi_{ij}$ characterizes completely the process and $\sigma_i$ are the Pauli matrices $\sigma_0=\id,\ \sigma_1=\sigma_X,\ \sigma_2=\sigma_Y,\ \sigma_3=\sigma_Z$. We used as benchmarks three relevant cases pertaining to a single-parameter qubit channel, a two-parameter qubit channel, and a single-parameter two-qubit channel. These represent relevant tests due to the matrices characterizing the channels but also because they are related to real-life scenarios.

{\it Depolarizing Channel (DC)-} A DC transforms an input qubit state $\rho$ as
$\Gamma_{\{p\}}[\rho]= {\sum^3}_{i=0}p_i \sigma_i  \rho \sigma_i$ with $p_0=1-p, p_1=p_2=p_3=p/3$. 
The associated process matrix $\chi$ acting on one qubit of a Bell state, e.g. $\ket{\phi^{+}}=\frac{1}{\sqrt{2}}(\ket{00}+\ket{11})$, reads as follows:
\beq\label{DC}
\begin{aligned}
\chi_{DC}&=\Gamma_{\{p\}}\otimes \id \left[\ket{\phi^{+}}\bra{\phi^{+}}\right]\\
&=\begin{pmatrix}
1-p & 0 & 0 & 0\\
0 & p/3 & 0 & 0\\
0 & 0 & p/3 & 0 \\
0 & 0 & 0 & p/3
\end{pmatrix},
\end{aligned}
\eeq
according to the expression in the Pauli basis. The DC serves as a useful model for imperfections in communications, thus our work can find application in the context of quantum channel estimation \cite{PhysRevA.63.042304}
.

{\it Generalized Qubit Amplitude Damping (GAD)-} The GAD, one of the sources of noise in superconducting-circuit-based quantum computing, is a two parameter family of channels where the parameters $\eta$ and $\gamma$ are between zero and one. 
The channel has a qubit input and qubit output and its superoperator has the form:
\beq\label{GAD}
\begin{aligned}
\chi_{GAD}&=\Gamma_{\{\eta,\gamma\}}\otimes \id \left[\ket{\phi^{+}}\bra{\phi^{+}}\right]\\
\Gamma_{\{\eta,\gamma\}}[\rho]&= {\sum^3}_{i=0} K_i  \rho K^{\dagger}_i,
\end{aligned}
\eeq
where $K_0= \sqrt{1-\gamma} ( \ket{0}\bra{0} +  \sqrt{1-\eta}\ket{1}\bra{1})$, 
$K_1= \sqrt{\eta(1-\gamma)} \ket{0}\bra{1} $, 
$K_2= \sqrt{\gamma} (\sqrt{1-\eta} \ket{0}\bra{0} + \ket{1}\bra{1}) $, and $K_3= \sqrt{\eta\gamma} \ket{1}\bra{0}$ are Kraus operators. The explicit equation of $\chi_{ij}$ is derived for each pair of $\{\eta,\gamma\}$ from \eqref{GAD}, but its explicit expression is not reported.

{\it Controlled-Phase (CP)-} Noisy quantum gates are fundamental for quantum technologies, {\it e.g.} quantum protocols for quantum computing and quantum information. Here we considered the C-Phase, an example of 2-qubit quantum gate.
The $\chi_{CP}$ process can be derived by the corresponding two-qubit unitary gate \cite{PhysRevApplied.13.034013,White:07}:
\beq\label{CP}
U_\phi =\begin{pmatrix}
1 & 0 & 0 & 0\\
0 & 1 & 0 & 0\\
0 & 0 & 1 & 0 \\
0 & 0 & 0 & e^{-i \phi}
\end{pmatrix}
\eeq
Similarly to the case of the GAD, the explicit expression of $\chi_{ij}$ is derived from the action of $U_\phi$ for each value of $\phi$ but we do not report it explicitly.

\subsection{NNs architecture, parameters and training}
\begin{figure*}[h]
\begin{center}
\includegraphics[width=\textwidth]{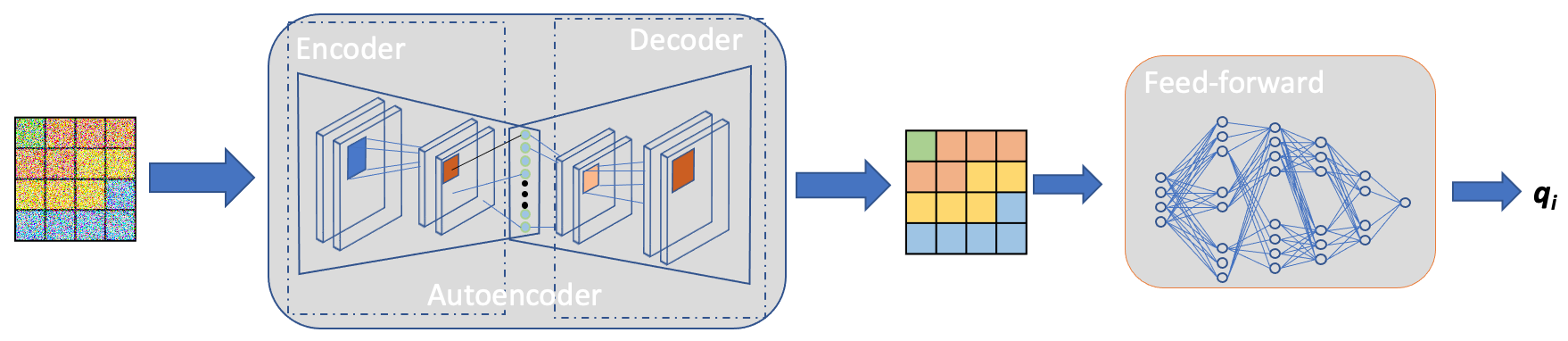}
\caption{The scheme shows the process employed for the characterization of the sought parameters $q_i$ and consists of two main steps: (a) simulated noisy process matrix are fed to the autoencoder for denoising; (b) the denoised matrices are used as input for the feed forward network, which estimates the parameters $q_i$ characterizing the quantum channel.}
\label{fig_pipeline}
\end{center}
\end{figure*}

Our parameter extraction technique is based mainly on an architecture composed of a convolutional NN (CNN), i.e. an autoencoder NN (aNN), and a dense (or fully) connected feed-forward NN (FFNN). The first attempts to remove noise in the process matrix obtained by the simulated data for the quantum channels, the second infers the parameters $q_i$ characterizing the channel itself. This architecture is similar for all the three channels considered in this manuscript and the specific layers' composition was suitably adapted to each case by choosing empirically the best set of hyperparameters.
A schematic depiction of the entire workflow is represented in Fig. \ref{fig_pipeline}. The idea to use specifically these two architectures comes from image denoising processing and classification.

We feed the aNN with the noisy complex process matrix, which the network interprets as a two-channel gray-level image. The purpose of the aNN is to populate the latent space only with the relevant components extracted with the 2D convolutional layers of the encoder, and then reconstruct the complex matrix processing the components in the latent space with the transposed 2D convolutional layers of the decoder \cite{goodfellow16,Hashisho2019}. 
All the Networks used in this work were trained with the Minimum Squared Error (MSE) loss function, given by the equation:
\begin{equation}
\mathcal{L}_{\text{MSE}} = \frac{1}{n} \sum_{i=1}^{n} (y_i - \hat{y}_i)^2
\end{equation}
where $n$ is the number of samples in a single batch, $y_i$ is the expected output value, and $\hat{y}_i$ is the predicted value of the parameters ({\it e.g.} $p$ for the DC or $\phi$ for the C-Phase). The choice of the MSE loss is particularly indicated in such regression problems, where the predicted quantities take continuous values.



\begin{figure*}[h]
\begin{center}
\includegraphics[width=\textwidth]{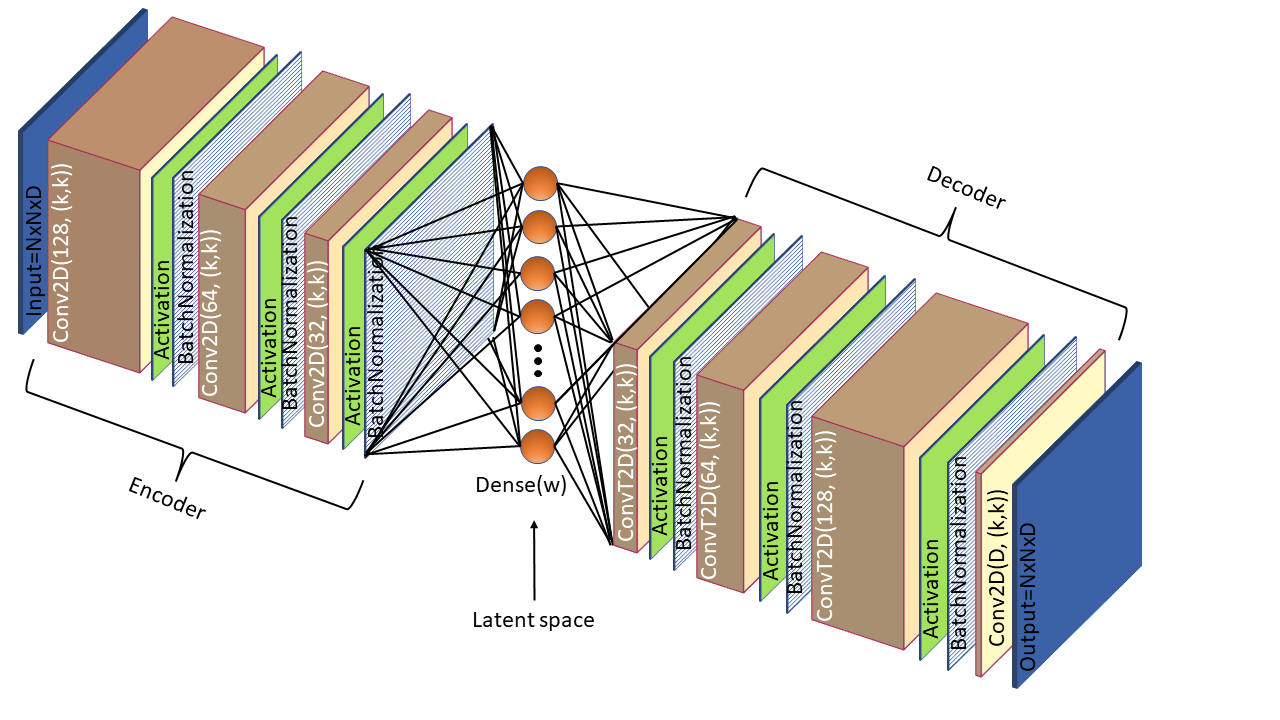}
\caption{General Autoencoder architecture used for denoising the three quantum channels. The kernel size (k) and the latent space shape (w) were determined empirically}
\label{fig_autoencoder}
\end{center}
\end{figure*}

{\it The Autoeconder NN -} The aNN architecture is composed of two main instances, called respectively encoder and decoder (Fig. \ref{fig_autoencoder}). Both of them are composed of three convolutional (encoder) and deconvolutional (decoder) layers, followed by a rectified linear unit activation layer, which introduces non-linearity in the model avoiding the problem of the vanishing gradient, and a batch normalization layer, to stabilize the learning process and reduce the number of training epochs. Between the encoder and decoder layers there is a dense layer (composed of single neuron units fully connected), which determines the so-called latent space, a representation of compressed data where similar data points are closer together in space: in our case the single quantum channel features without the simulated noise. Even if the latent space is not shown in Fig. \ref{fig_autoencoder} for graphical needs, it is fully connected by flattening the output of the last batch normalisation layer of the encoder and reshaping the output from the latent space so to be compatible with the input of the first layer of the decoder. Tab. \ref{Table_autoencoder_param} shows respectively the kernel size of each convolutional stage and the dense layer shape for the quantum channels under investigation. 

\begin{table}[h]
\caption{kernel size (k) and latent space shape (w) for each channel}
\centering
\begin{tabular}{c | c | c | c }
\hline
\hline
\multicolumn{4}{c}{aNN} \\
\hline
\hline
 & DC &  GAD &  CP \\
\hline
k & 2 &  2 & 4 \\
\hline
w & 40 &  40 & 30\\
\hline

\end{tabular}
\label{Table_autoencoder_param}
\end{table}

\begin{figure*}[h]
\begin{center}
\includegraphics[width=\textwidth]{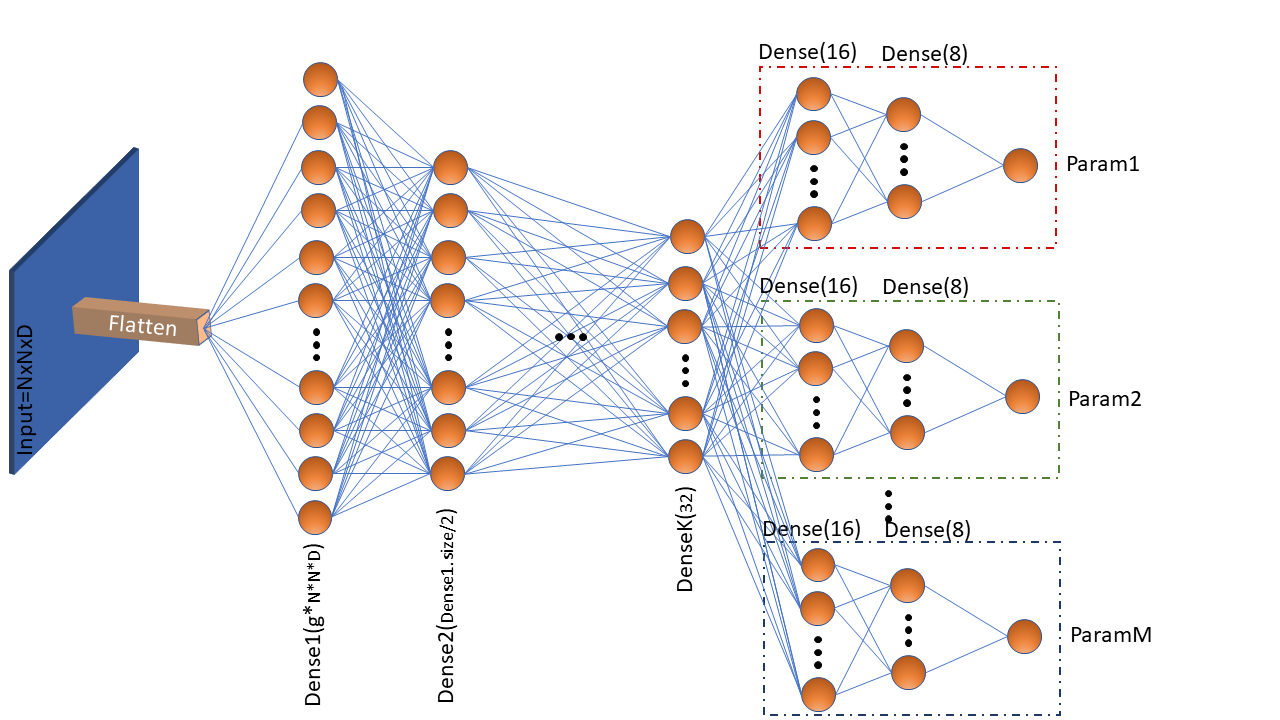}
\caption{The scheme shows the process employed for the characterization of the sought parameters $q_i$ and consists of two main steps: (a) simulated noisy process matrix are fed to the autoencoder for denoising; (b) the denoised matrices are used as input for the feed forward network, which estimates the parameters $q_i$ characterizing the quantum channel.}
\label{fig_ffnn}
\end{center}
\end{figure*}

{\it The Feed-Forward NN -} 
The input layer is represented by the denoised complex matrices, i.e. the output of the aNN. The first layer of this network is composed of a number of neurons equal to the size of the flattened array, obtained from the input matrices, multiplied by a factor $g$, as shown in Tab. \ref{Table_ffnn_param}. This factor was determined empirically and increases as the complex matrices shapes decrease. This NN is also characterized by the presence of a number of two-layer branches equal to the number of the parameters to be estimated, as shown in Fig. \ref{fig_ffnn}. 


\begin{table}[h]
\caption{FFNN architecture parameters}
\centering
\begin{tabular}{c | c | c | c }
\hline
\hline
\multicolumn{4}{c}{FFNN} \\
\hline
\hline
 & DC &  GAD &  CP \\
\hline
g & 2 &  2 & 1 \\
\hline
branch & 1 & 2 & 1\\

\hline

\end{tabular}
\label{Table_ffnn_param}
\end{table}


{\it NN training -} The training of the whole NN includes hundreds of process matrices $\tilde\chi^{(\alpha)}$,  obtained from simulated process tomography data with a population of $n_i$ events on average, characterized by a Poissonian distribution; we have then considered three signal levels $n_i=k_i n$ with $n=2000$, $k_i=0.1, 0.5, 1$. Both blocks of the NN are trained using back-propagation, adopting the mean squared error (MSE) as loss function. The dataset is split between training and validation, with a $0.8-0.2$ ratio. The two-block NN reaches a MSE as low as ($\approx 10^{-4}$) after $\approx 8$ epochs during validation, which can be further improved by increasing the number of epochs. More details on the generation of the dataset are contained in the Appendix.

\section{Results}

The tests we carried out have inspected the following cases:
\begin{itemize}
    \item depolarising channel (DC), described by a single parameter $p$. We consider the range $p \in [0.05,1]$, with 100 instances for the training for each value;
    \item generalised amplitude damping (GAD), described by two parameters $\{\eta,\gamma\}$. We consider the range $[0,1]$ in steps of 0.1, with 500 instances for the training for each value;
    \item controlled-phase two-qubit gate (CP), described by a single parameter $\phi$. We consider $\phi= m_1 \cdot \pi/6$ with $m_1=[0,11]$ and $\phi= m_2 \cdot \pi/4$ with $m_2=1,3,5,7$, with 500 instances for the training for each value.
\end{itemize}

\subsection{Depolarising channel}

As a first test, we consider a DC and the case of data only being affected by Poissonian noise associated to the simulated measured counts. Since we are concerned with the accuracy, we considered as the relevant figure of merit the residue $\vert p_{\rm meas}-p_{\rm set}\vert$ between the inferred  $p_{\rm meas}$ and the set $p_{\rm set}$ values of $p$ for three different approaches: our optimised network; a simpler network only adopting the feed-forward (FF) block; the standard evaluation taking the highest fidelity with a state in the form $ \Gamma_{\{p\}}[\rho]= (1-p)\rho+p/3\sum^3_{i=1}{\sigma_i \rho \sigma_i}$, what we term Maximum Fidelity (MF). Fidelity is defined as:

\begin{equation}
F \propto Tr[\sqrt{\sqrt{\chi_{exp}} \chi_{teo} \sqrt{\chi_{exp}} }]^2
\end{equation}
where $\chi_{exp}$ is the process matrix obtained with the process tomography applied on the experimental data and $\chi_{teo}$ is the theoretical one. The parameter can be estimated maximizing the value of $F$ as a function of p.

\begin{figure}[h]
\begin{center}
\includegraphics[width=0.67\textwidth]{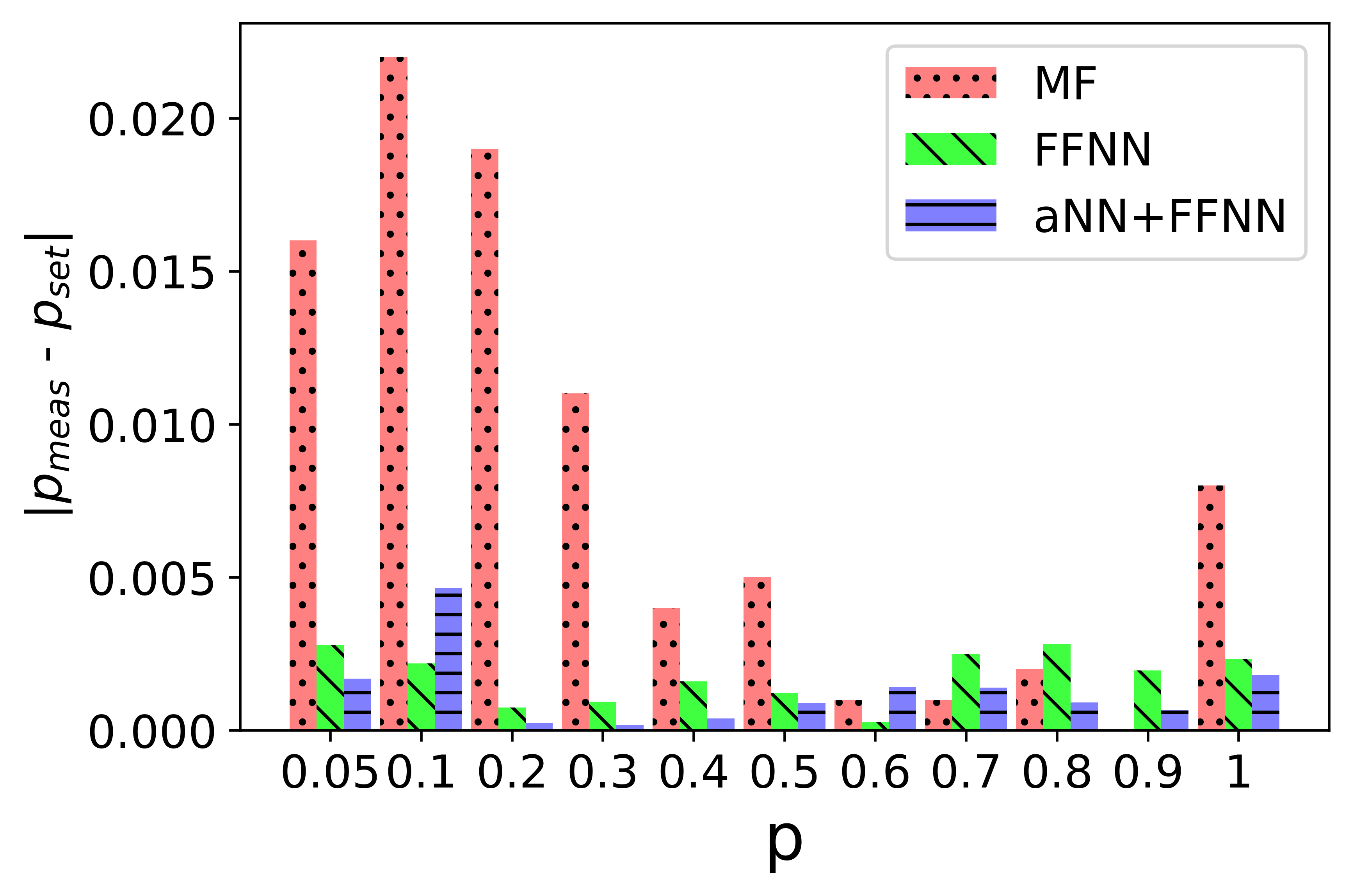}
\caption{Mean values of the residues, i.e. $|p_{meas}- p_{set}|$, obtained with three different evaluation strategies: employing the optimized two-stage NN (blue, horizontal lines), a simple FFNN (green, diagonal lines) and evaluating the MF (red, dotted). The test of the NNs is performed using 300 simulated process matrix for each value of $p$.}
\label{residues}
\end{center}
\end{figure}

The summary in Fig.~\ref{residues} shows the appraisal of including the convolutional aNN stage with respect the only FFNN. This demonstrates that our two-stage network can outperform the simple MF and offers some advantage over the use of the FF only.

The main advantage of our approach unfolds when considering instances with parasitic processes. For this second test case, we employed our NN to data generated in a real experiment~\cite{chiuri11prl} performed with an all-optical setup, where we expect deviations from the ideal depolarising channel. As before, the training was performed with simulated data from the \emph{ideal} process. This is aimed at accounting for the expected fluctuations in the data, with an ideal output, without incorporating knowledge on the actual process. This implies that the matrix serving as the input of the FF network should not be interpreted as the experimental $\chi$ matrix. This is rather the representation of the closest DC to the data.

\begin{figure*}[h]
\begin{center}
\includegraphics[width=0.8\textwidth]{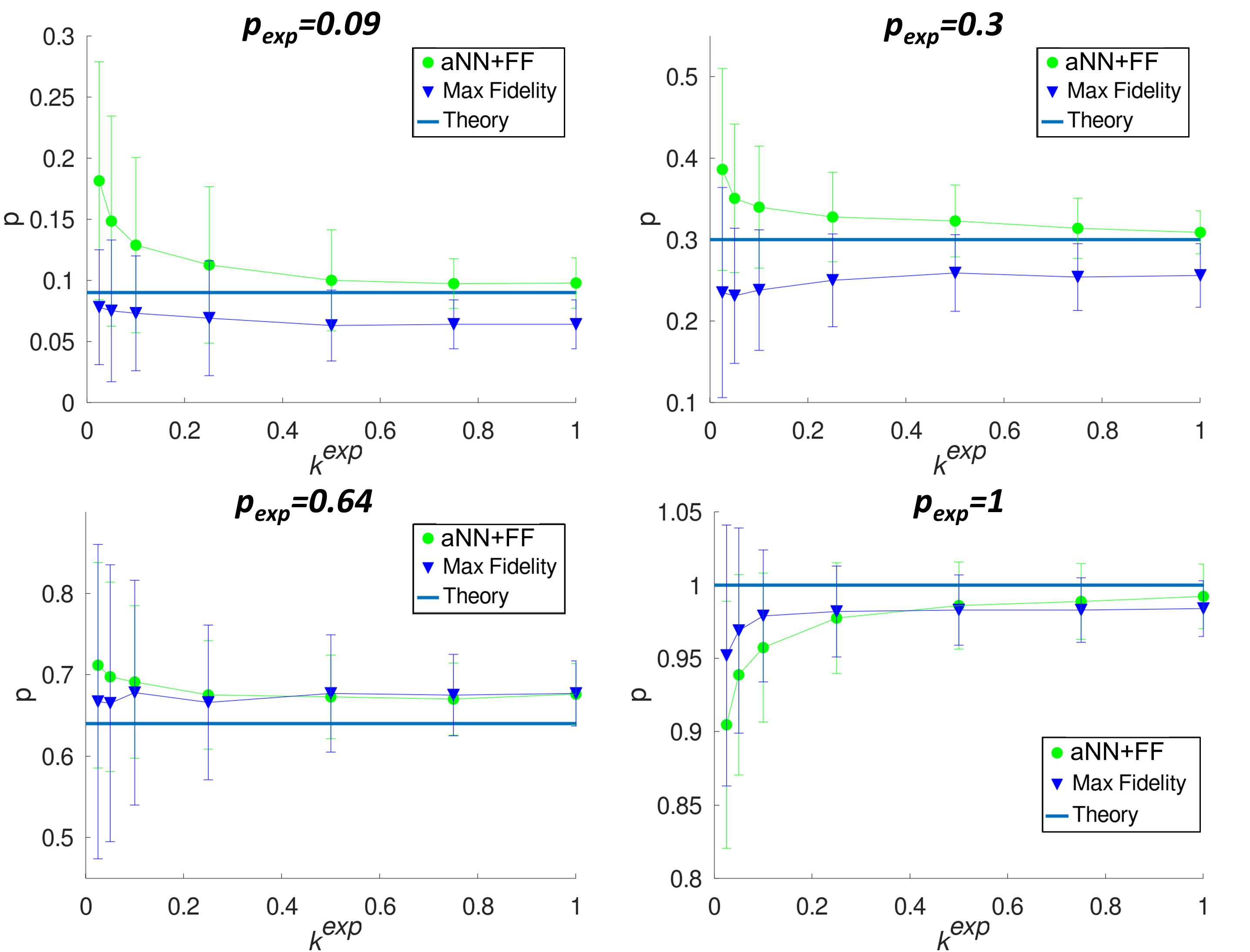}
\caption{Evaluated $p$ from the experimental data set for different expected values $p_{exp}$ as a function of the employed resources. The results show the expected value (blue line) and the results obtained through our method (green circles) and through maximum fidelity (blue triangles).Errors are obtained through standard deviation.}
\label{results_estim}
\end{center}
\end{figure*}

The data were collected for the expected values $p_{\rm exp}=0.09,0.3,0.64,1$ determined by varying the activation time of the Pauli operators experimentally implemented, as described in detail in the Appendix. In order to investigate the impact of the Poissonian noise on the reconstruction we then generated 300 matrices from the data corresponding to each $p_{exp}$ for each rescaling factor $k^{\rm exp}_i=0.025,0.05,0.1,0.25,0.5,0.75,1$ on the overall average number of measured counts $n^{\rm exp}$, {\it i.e.}  $n^{exp}_i= k^{exp}_i  n^{exp}$  with $n^{\rm exp}_7 \approx 2000$ comparable with $n_3$. The results obtained with the experimental dataset are reported in Fig.\ref{results_estim}. There, we plot the estimated values of $p$ as a function of the rescaling factor for our four cases using both our NNs, and a simple procedure based on maximising the fidelity. Overall, we assess a satisfactory behaviour in the evaluation of $p$. The best performances is achieved for small values of $p$ (i.e. $p=0.09,0.3$), representing the most interesting cases, in that the noise can easily overcome the channel: the adoption of the NN analysis seems to be able to mitigate biases, and this holds true down to $k^{exp}_i=0.5$ for $p=0.09,0.64,1$ and $k^{exp}_i=0.25$ for $p=0.3$, showing that the presence of parasitic processes does not introduce costs in terms of resources needed. For higher values of $p$ (i.e. $p=0.64,1$), instead, the performance of the two methods is aligned, with marginal amelioration still appreciable for $k^{exp}_i \geq 0.5$. This difference can be traced to the form of the process: for low $p$, uncertainties due to the statistics are more relevant, as diagonal terms in the matrix are smaller. The experimental conditions when $p=0.64$ were not optimal and the  parameter that was actually observed is slightly higher; this shows how implemented channel depolarises more than expected. This affects the measured counts and the process matrices reconstructed with the two methods reported in Fig.\ref{results_estim}. Nevertheless, our approach allowed us to obtain a matrix very similar to the expected one and the effects due to experimental imperfections seems to be contained for almost all values of $p$.

\subsection{Generalised amplitude damping channel and control-phase gate} We now consider different, more involved examples of quantum channels. The first example concerns the GAD, characterised by a damping parameter $\gamma$, and by the probability $\eta$ of dissipating towards either level of the qubit. As a more compact figure to assess the effectiveness of our cascaded network, we adopt the success probability obtained as follows: we consider samples composed by $\approx 100$ instances for each pair $\{\eta,\gamma\}$. For each sample, we evaluate the percentage of success, i.e. the obtained parameter $q_i$ in the range $|q_i|\leq 0.1$, and these values are used to estimate how many times the application of the proposed approach allows to reach a desired threshold, {\it e.g.} $>99\%$. This procedure was repeated for different rescaling factors $k_i$ and the results are summarised in Table~\ref{Table_GAD6}, presenting typical occurrences. Even for the lowest signal level, $k_1=0.1$, introducing the aNN it is possible to achieve superior performance with respect to the simple application of a FFNN, thus demonstrating the usefulness of the proposed double-stage architecture.


\begin{table}[h]
\caption{Summary of the results for GAD. For each pair $\{\eta,\gamma\}$ we tested our approach over a sample of  $\approx 100$ matrices. We report the success rate to obtain a residue $\leq0.1$ for $>99\%$ of the testing sample.}
\centering
\begin{tabular}{c | c c | c c | c c}
\hline
\hline
 & \multicolumn{2}{c}{$k_1=0.1$} &  \multicolumn{2}{c}{$k_2=0.5$} &  \multicolumn{2}{c}{$k_3=1$}\\
\hline
Parameter & FF & aNN+FF &  FF & aNN+FF &  FF & aNN+FF\\
\hline
\hline
$\eta$ & 0 $\%$ &  37  $\%$ & 13.6 $\%$ & 75.3  $\%$ & 34.6 $\%$ & 84  $\%$\\
$\gamma$ & 0 $\%$ &  16$\%$  &  42$\%$ &  100$\%$ &  88.8$\%$ &  100$\%$\\
\hline
\hline
\end{tabular}
\label{Table_GAD6}
\end{table}

We have adopted a similar approach to test the performance on a CP gate. The figure of merit is, however, slightly modified so to make it more appropriate to the case in point: we consider the probability of achieving an error permitting a full discrimination of all our angles. Even at the lowest signal level, almost all events show such discrimination ability, with a significant improvement over the use of FFNN only [See Table~\ref{Table_cphase}]. This is also indirectly inferred by comparing the matrices obtained after the aNN with those obtained with the usual maximum likelihood routine. We observe how aNNs achieve better performance in that the fidelity with the expected matrix is always $\geq 0.99$, with a significant fraction of values being  $1$, even at the lowest signal, $k_1=0.1$. With usual maximum likelihood method, instead, the fidelity generally remains below $ 0.99$.

\begin{table}[h]
\caption{Summary of the results for CP. For each value of $\phi$, we tested our approach over a sample of 500 matrices. We report the success rate for both the absolute and the relative residue ($\leq 3\%$).}
\centering
\begin{tabular}{c | c c | c c | c c}
\hline
\hline
 & \multicolumn{2}{c}{$k_1=0.1$} &  \multicolumn{2}{c}{$k_2=0.5$} &  \multicolumn{2}{c}{$k_3=1$}\\
\hline
Residues & FF & aNN+FF &  FF & aNN+FF &  FF & aNN+FF\\
\hline
\hline
$\phi$ (Abs) & 4.7 $\%$ & 99.8  $\%$  & 10.5 $\%$ & 100 $\%$ & 11.1 $\%$ & 100  $\%$\\
$\phi$ ($\%$) & 3.4 $\%$ &  92.6 $\%$  & 6.6 $\%$ & 96.3 $\%$ & 6.7 $\%$ & 98.5 $\%$\\
\hline
\hline
\end{tabular}
\label{Table_cphase}
\end{table}

More detail on the analysis of GAD and CP are found in the Appendix.

\section{Discussion}
In this work we have presented a model unveiling the potential of an optimized architecture based on NNs to achieve characterization of quantum objects from noisy measurements: the relation between the observed frequencies and underlying parameters of the measured process were approximated by a two-stage NNs. We applied the presented approach to both numerical and experimental dataset, demonstrating that the employment of these tools tolerates at least a $50\%$ reduction in the number of collected counts even under unfavourable conditions. A key result is the adoption solely of simulated data for the training. Despite this, our NNs architecture has correctly identified the new scenarios set by $p_{exp}$ and $k^{exp}_i$.
This spares the need of collecting a large amount of data, and ensures that the network is trained on the Poissonian noise, and not on spurious effects taking place in the experiments.
It is then possible to guarantee stability and reliability, while minimizing the resources and time necessary to train the NNs.

The NN approach can then provide accurate extraction of the parameters, with a more satisfactory performance than the direct use of estimates based on maximum likelihood methods. The computational resources, however, do not result heavily aggravated as a consequence. On the other hand, Monte Carlo routines become necessary for the estimation of error, due to the nontrivial correlations employed in the network, especially in its convolutional layers. This lays a promising route for routine employment in more complex instances. Indeed, our tests have addressed different quantum channels, and the two-stage NN have been optimized for the particular process. However, they covered a broad class, making us confident that the result obtained with this demonstration may have general validity. Our approach then allows to characterize the process, reduce the impact of the noise, and extract the parameters $q_i$, while optimizing the amount of necessary resources for real-life scenarios. We have illustrated applications directly linked to quantum communications and quantum computing: DC is a standard description of communication channels, while GAD is often employed to model decoherence processes in atomic, superconducting and solid-state qubits. At the merging point, distributed computing, embracing both communication and calculation tasks, may take particular advantage from our procedure. Extensions to higher-dimensional systems will also benefit from the reduced requirements in terms of data acquisition.

\subsection*{Acknowledgments}
The authors thank P. Mataloni for granting access to the data and C. Macchiavello for fruitful discussion. This work was supported by the European Commission (FET-OPEN-RIA STORMYTUNE, Grant Agreement No. 899587), and by the NATO SPS Project HADES - MYP G5839.

\subsection*{Authors contribution}
A.C. conceived and supervised the project, and carried out the analysis of the experimental data. I.G. and M.B. helped refining the focus of the investigation. M.G. implemented the neural network algorithms. All authors contributed to the discussion, analysis of the results and the writing of the manuscript.

\subsection*{Competing Interests}
The authors declare that there are no competing interests.

\subsection*{Data Availability}
The data that support the findings of this study are available from the corresponding author upon reasonable request.

\section*{Appendix}


\subsection*{Experimental Process Tomography} 
We briefly discuss the setup employed in Ref.~\cite{chiuri11prl}. This is based on a two-photon source producing entangled states in polarisation.

\begin{figure}[h!!]
\begin{center}
\includegraphics[width=0.5\columnwidth]{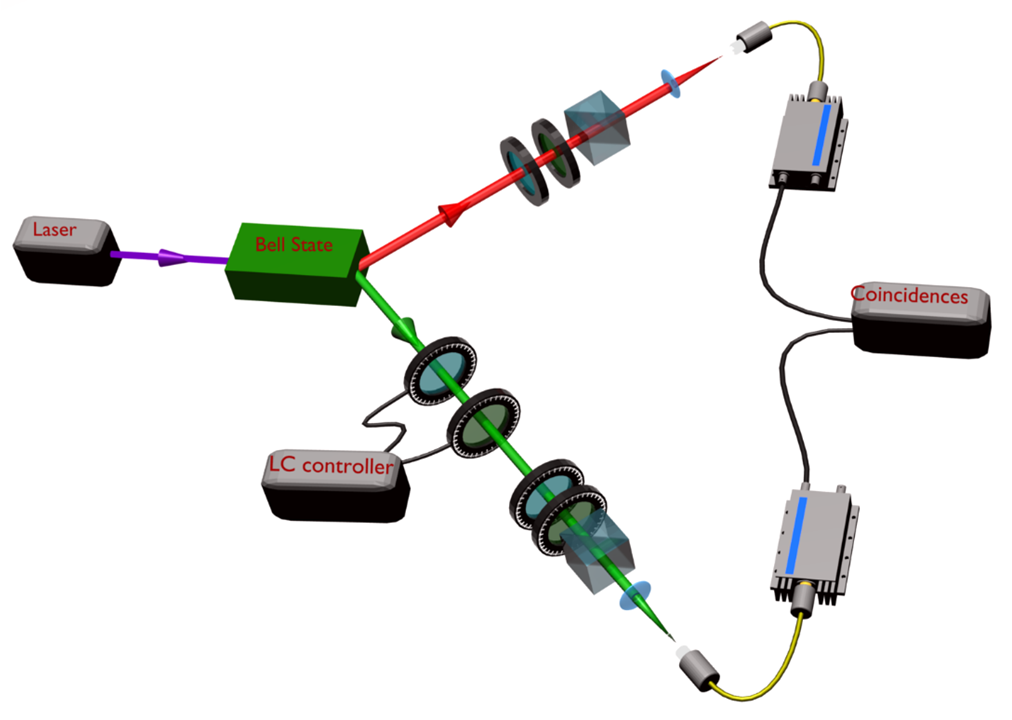}
\caption{Experimental Setup. The qubit encoded in the green photon is affected by a DC, implemented by suitably activating two LCs. We generated four different channels by varying the parameter $p_{exp}=0.09,0.3,0.64,1$ by controlling their relative delays and activation times. State tomography was realized by the usual arrangement of polarization analysers placed in front of each detector.}
\label{setup}
\end{center}
\end{figure}

The general Pauli channel (PC) is implemented by means of a sequence of liquid crystal retarders (LC1 and LC2) in the path of one of the photons -- see Fig.~\ref{setup}.
The LCs act as phase retarders, introducing a relative phase between the ordinary and extraordinary polarisation components depending on the applied voltage $V$. Precisely, $V_{\pi}$ and $V_{\id}$ correspond to the
case of LCs operating as a half-waveplate (HWP) and as the
identity operator, respectively. The LC1 and LC2 optical axes are set at $0^\circ$ and $45^\circ$ with respect to the $V$-polarization.
When the voltage $V_{\pi}$ is applied, the LC1 (LC2) acts as a Pauli $\sigma_z$ ($\sigma_x$) on the single qubit. We were able to switch between $V_{\id}$ and $V_\pi$ in a controlled way and independently for both LC1 and LC2. The simultaneous application of $V_{\pi}$ on both LC1 and LC2 corresponds to the $\sigma_y $ operation. We could also adjust the temporal delay between the intervals in which the $V_\pi$ voltage is applied to the two retarders. We define $t_1$, $t_2$, $t_3$ respectively as the activation time of the operators $\sigma_x$, $\sigma_y $ or $\sigma_z$ and $T$  is the period of the LCs activation cycle. 

A general PC, i.e. with anisotropic noise, was generated by varying the four time intervals $t_1$, $t_2$, $t_3$ and $T$. These intervals $t_i$ are related to the probabilities $p_i$ ($i=1,2,3$) by the following expression: $p_i=\frac{t_i}{T}$.
The probability $p_0$ of the identity operator is given by $p_0=1-\frac\delta T$ (with $\delta=t_1+t_2+t_3$).
The condition $t_1=t_2=t_3$ corresponds to the depolarizing channel, i.e. i.e. isotropic noise, with the three Pauli operators acting on the single qubit with the same probability $p=\frac{\delta}{T}=\frac{t_1+t_2+t_3}{T}$.
In the experiment, the parameter $p_{exp}$ was varied by changing the interval $T$ for a fixed period $\delta$.

The reconstruction of the quantum process matrices $\chi_{ij}$ is carried out by means of ancilla-assisted quantum process tomography~\cite{Alteprl03,mohseni08pra}. This follows the procedure: i) preparing a two-qubit maximally entangled state ii) sending one of the two entangled qubits through the channel $\mathcal E$; iii) reconstructing the output two-qubit state by state tomography ~\cite{James01}. The density matrix $\chi$ from the two-qubit output density matrix is the process matrix. 

\begin{figure}[h]
\begin{center}
\centerline{\includegraphics[width=0.7\columnwidth]{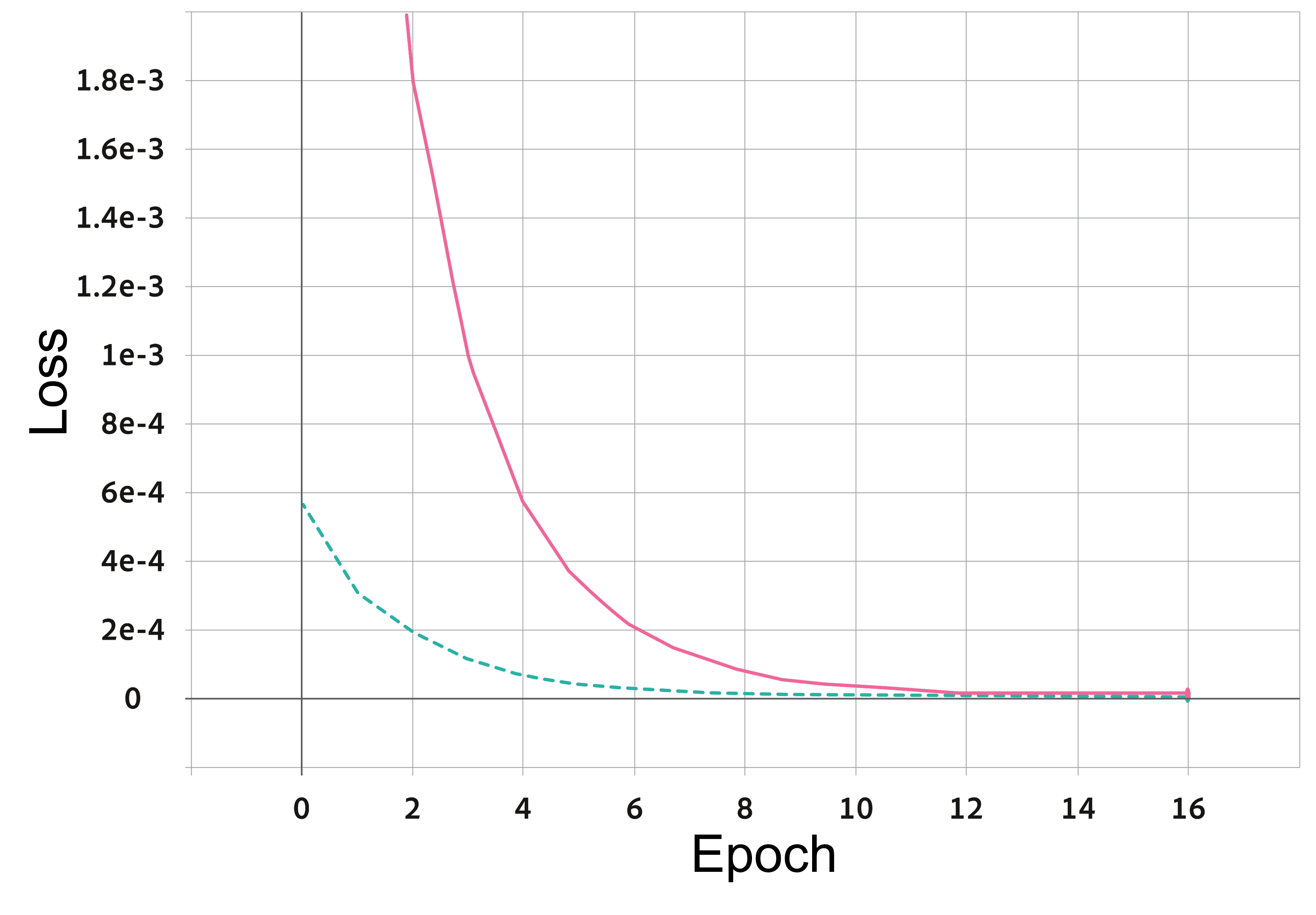}}
\vspace*{8pt}
\caption{Training (red - solid) and validation (green - dashed) loss curves for the aNN: after 12 epochs the trained aNN is able to remove the noise from the data input. They are correctly decreasing and the validation curve is always under the training one as expected when the NN training is properly performed.}
\label{fig:loss_vs_epochs}
\end{center}
\end{figure}

\subsection*{Details on the networks}

During the training and validation phase of a NN it is always essential to avoid the overfitting of the network. In this work different strategies were adopted for avoiding the overfitting problem during the training phase of the aNN. 

Good results were obtained testing two different configurations: training using the simulated matrices (i.e. implemented for GAD and CP), training based on a dataset composed of mixed matrices (i.e. implemented for DC).

\begin{figure}[h]
\includegraphics[width=\columnwidth]{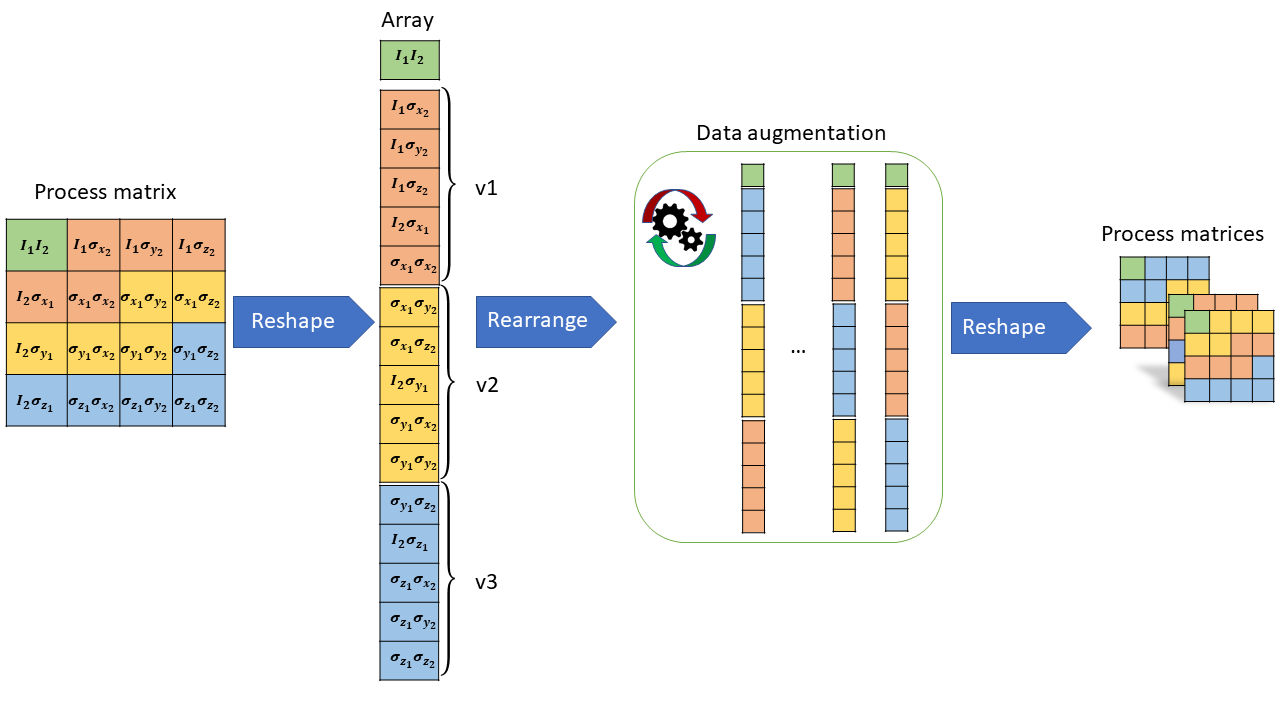}
\caption{In order to increase the size of the input data set for the DC, the simulated process matrices used to train the aNN have been rearranged in different combinations. Each matrix was divided in four arrays: a $1\times 1$ array (green) containing the first element, and three $5\times 1$ arrays (orange, yellow, and blue) each including four off-diagonal elements and a diagonal one. These four arrays were then reshaped as a $16 \times 1$ array and rearranged in $5$ different combinations, having the $1\times 1$ array, corresponding to $1-p$ value always as first, while reshuffling the others. The $16 \times 1$ was then mapped back to a $4 \times 4$ matrix and fed to the aNN.}
\label{fig_augmentation}
\end{figure}

Indeed, in this case we took advantage of the symmetry of the matrix \eqref{DC} which was reshaped taking the elements in lexicographic order as a 16-element vector. The latter was then divided in four sub-arrays with dimensions $1\oplus 5 \oplus 5 \oplus 5$ and the last three sub-arrays were exchanged in five different combinations to give valid matrices [See Fig.~\ref{fig_augmentation}]. Hence, we demonstrated that in this particular case it is possible to exploit an artificial dataset augmentation to train the NNs employing less resources and time. The GAD and the CP are not characterized by useful symmetries and a suitable number of random matrices was generated to compose the training data set. The absence of the overfitting was ensured by the loss function trend on each epoch [See Fig.\ref{fig:loss_vs_epochs}]. 

The properties characterizing the DC were exploited also to further simplify the parameters estimation. Indeed in this case only the four diagonal terms were considered as input of the FFNN. On the other hand, we have proved that our results could be generalized and we tested our approach to the GAD and CP channels to demonstrate that the proposed approach works also in non-trivial scenarios where the process matrix is not characterized by any symmetry and its elements are not directly related to the parameters.

\subsection*{Additional numerical analysis}
\begin{figure*}[h!!]
\includegraphics[width=\textwidth]{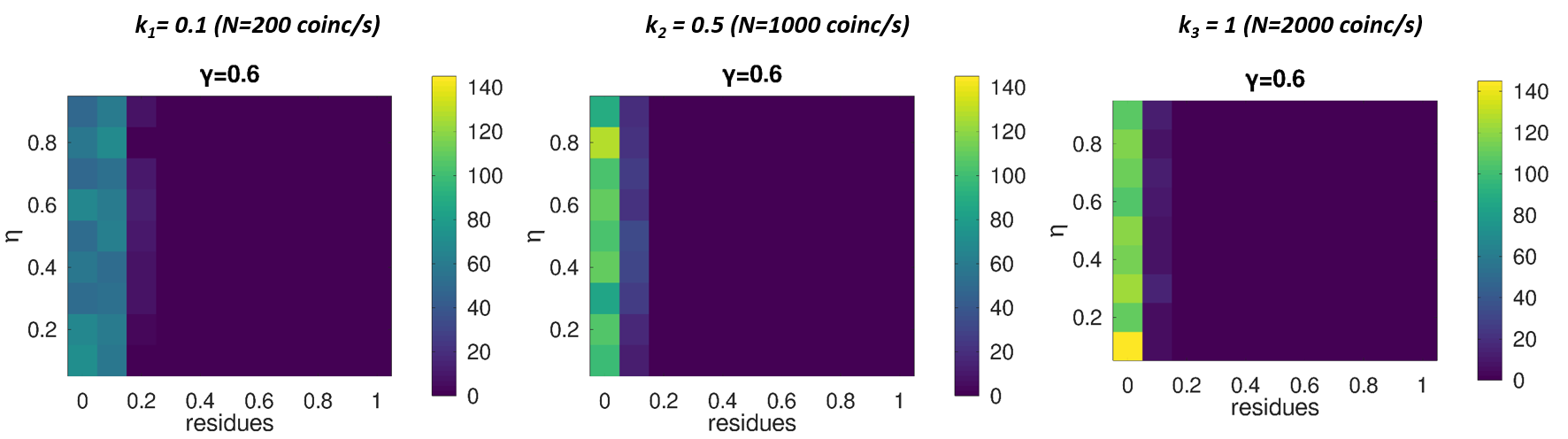}
\caption{GAD: even for the lowest signal level, $k_1=0.1$, our combined NNs achieve a success rate $>90 \%$, i.e. probability of obtaining residues for the parameters below 0.1. We report here an example of what was obtained for all the considered values of parameters. The color map represents the number of test matrices for the particular value of residue.}
\label{GAD_rescal}
\end{figure*}

\begin{figure*}[h]
\includegraphics[width=\textwidth]{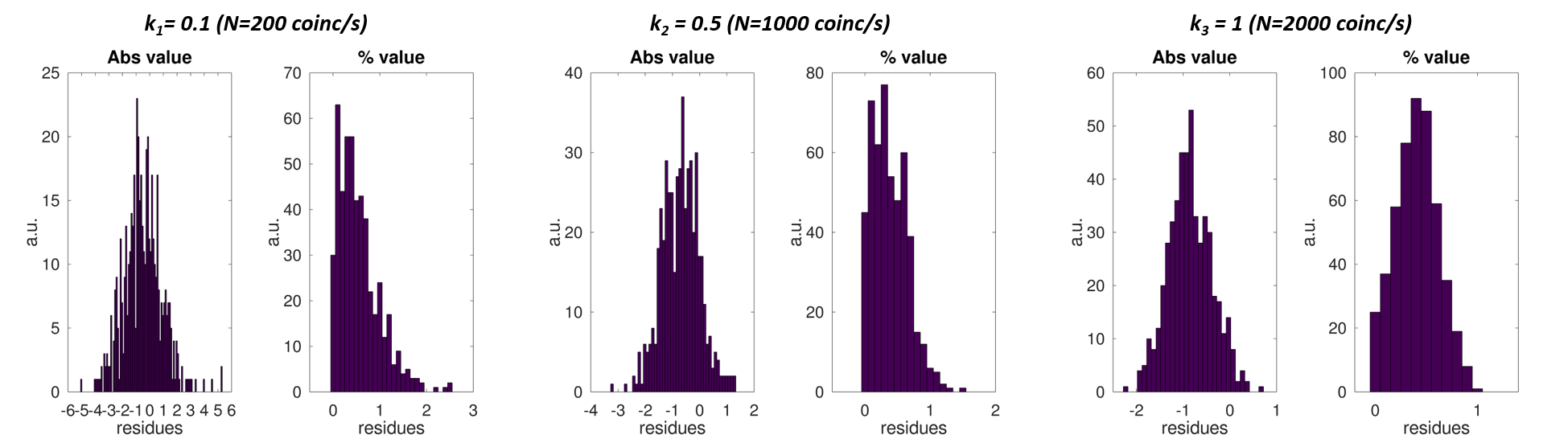}
\caption{CP ($\phi=210^{\circ}$): even for the lowest signal level, $k_1=0.1$, our combined NNs achieve a success rate $\approx 100 \%$, i.e. the probability of achieving an error permitting a full discrimination of all our angles. We report here an example of the results obtained for all the considered values of $\phi$. For each signal level $k_i$ we show two histograms: {\it left} {\it (right)} absolute (percent) value of the residues.}
\label{CPhase_rescal}
\end{figure*}

\begin{figure*}[h]
\includegraphics[width=\textwidth]{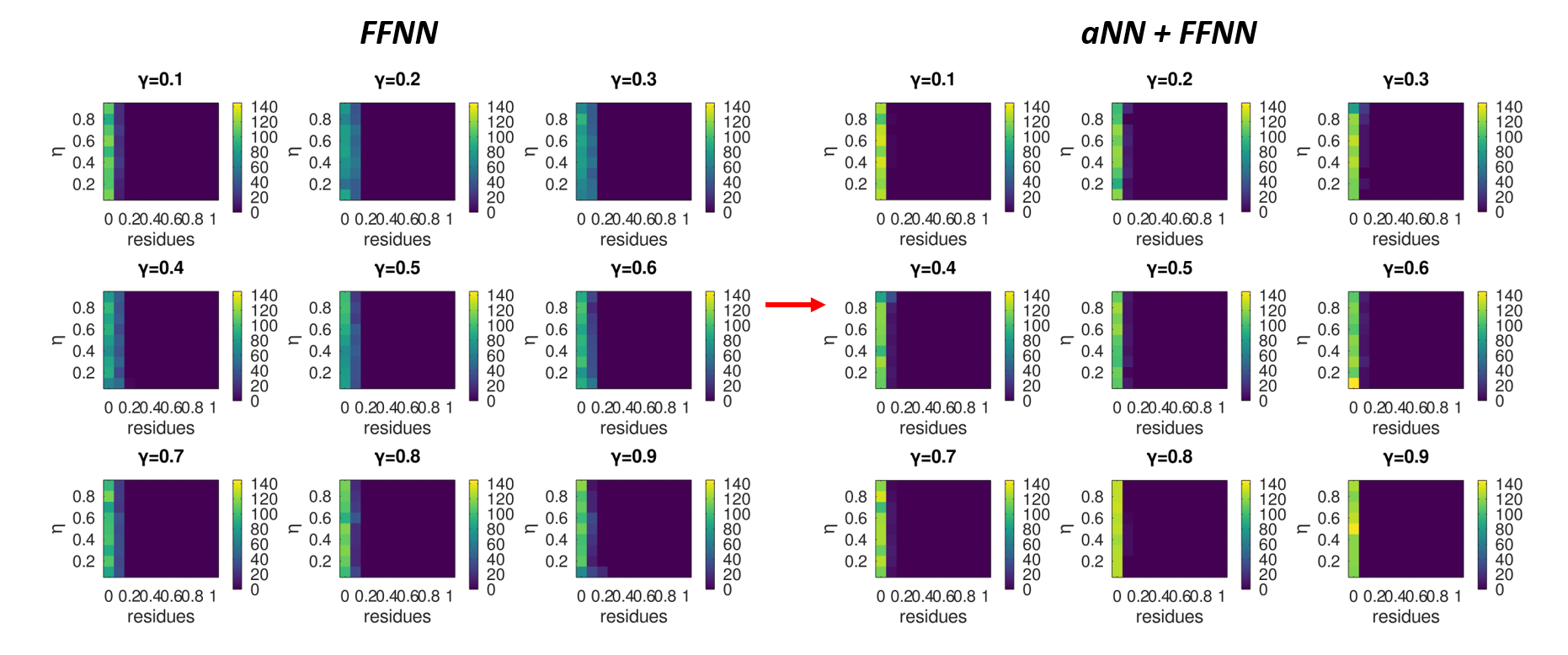}
\caption{GAD, $k_1=0.1$. The color map represents the number of test matrices for the particular value of residue. The employment of the aNN for the denoising allow to improve the success rate, i.e. to obtain a parameter closer to the expected one. }
\label{GAD_results}
\end{figure*}

We report here further results concerning the employed figures to assess the effectiveness of our cascaded network. For the GAD we adopt the success probability of obtaining residues for the parameters below 0.1 for at least $90\%$ (Table~\ref{Table_GAD5}) and $95\%$ (Table~\ref{Table_GAD7}) of the sample. The results are summarised also in Fig.~\ref{GAD_rescal} where it is confirmed that, even for the lowest signal level at $k_1=0.1$, our combined NNs achieve superior performance with respect to the simple application of a FFNN, thus demonstrating the usefulness of the autoencoder. The Fig.~\ref{CPhase_rescal} shows that similar results were achieved also for the CP. Fig.~\ref{GAD_results} demonstrates the fundamental contribution of the aNN allowing to feed the FFNN with denoised matrices.     

\begin{table}[h]
\caption{Summary of the results for GAD. For each pair $\{\eta,\gamma\}$ we tested our approach over a sample of $\approx 100$ noisy matrices. We report the success rate to obtain a residue $\leq0.1$ for $>90\%$ of the testing sample.}
\centering
\begin{tabular}{c | c c | c c | c c}
\hline
\hline
 & \multicolumn{2}{c}{$k_1=0.1$} &  \multicolumn{2}{c}{$k_2=0.5$} &  \multicolumn{2}{c}{$k_3=1$}\\
\hline
Parameter & FF & aNN+FF & FF & aNN+FF & FF & aNN+FF\\
\hline
\hline
$\eta$ & 2.5 $\%$ & 60.5  $\%$  &  44.4$\%$ & 87.7$\%$ &  61.7$\%$ &  90.1$\%$\\
$\gamma$ & 22.2$\%$ & 93.8 $\%$  &  96.3$\%$ & 100$\%$ &  98.8 $\%$ &  100$\%$\\
\hline
\hline
\end{tabular}
\label{Table_GAD5}
\end{table}

\begin{table}[h]
\caption{Summary of the results for GAD. For each pair $\{\eta,\gamma\}$ we tested our approach over a sample of  $\approx 100$ noisy matrices. We report the success rate to obtain a residue $\leq0.1$ for $>95\%$ of the testing sample.}
\centering
\begin{tabular}{c | c c | c c | c c}
\hline
\hline
 & \multicolumn{2}{c}{$k_1=0.1$} &  \multicolumn{2}{c}{$k_2=0.5$} &  \multicolumn{2}{c}{$k_3=1$}\\
\hline
Parameter & FF & aNN+FF & FF & aNN+FF & FF & aNN+FF\\
\hline
\hline
$\eta$ & 0 $\%$ &  46.9 $\%$  & 27.2  $\%$ &  81.5 $\%$ &  54.3$\%$ & 88.9 $\%$\\
$\gamma$ & 7.4 $\%$ & 40.7 $\%$ & 75.3 $\%$ & 100 $\%$ & 98.8 $\%$ & 100 $\%$\\
\hline
\hline
\end{tabular}
\label{Table_GAD7}
\end{table}

\bibliography{biblio.bib}

\begin{thebibliography}{10}
\providecommand \doibase [0]{http://dx.doi.org/}%

\bibitem{PhysRevA.66.062305}
G\"uhne O, Hyllus P, Bru\ss{} D, et al. Detection of entanglement with few
  local measurements. {\it Phys. Rev. A} 2002\string; 66\string: 062305.
\newblock \href {\doibase 10.1103/PhysRevA.66.062305} {doi:
  10.1103/PhysRevA.66.062305}

\bibitem{PhysRevLett.91.227901}
Barbieri M, De~Martini F, Di~Nepi G, Mataloni P, D'Ariano GM, Macchiavello C.
  Detection of Entanglement with Polarized Photons: Experimental Realization of
  an Entanglement Witness. {\it Phys. Rev. Lett.} 2003\string; 91\string:
  227901.
\newblock \href {\doibase 10.1103/PhysRevLett.91.227901} {doi:
  10.1103/PhysRevLett.91.227901}

\bibitem{PhysRevLett.92.087902}
Bourennane M, Eibl M, Kurtsiefer C, et al. Experimental Detection of
  Multipartite Entanglement using Witness Operators. {\it Phys. Rev. Lett.}
  2004\string; 92\string: 087902.
\newblock \href {\doibase 10.1103/PhysRevLett.92.087902} {doi:
  10.1103/PhysRevLett.92.087902}

\bibitem{Branderhorst_2009}
Branderhorst MPA, Nunn J, Walmsley IA, Kosut RL. Simplified quantum process
  tomography. {\it New Journal of Physics} 2009\string; 11(11)\string: 115010.
\newblock \href {\doibase 10.1088/1367-2630/11/11/115010} {doi:
  10.1088/1367-2630/11/11/115010}

\bibitem{Blandino_2012}
Blandino R, Ferreyrol F, Barbieri M, Grangier P, Tualle-Brouri R.
  Characterization of a $\pi$-phase shift quantum gate for coherent-state
  qubits. {\it New Journal of Physics} 2012\string; 14(1)\string: 013017.
\newblock \href {\doibase 10.1088/1367-2630/14/1/013017} {doi:
  10.1088/1367-2630/14/1/013017}

\bibitem{PhysRevA.84.050301}
Tipsmark A, Dong R, Laghaout A, Marek P, Jezek M, Andersen UL. Experimental
  demonstration of a Hadamard gate for coherent state qubits. {\it Phys. Rev.
  A} 2011\string; 84\string: 050301.
\newblock \href {\doibase 10.1103/PhysRevA.84.050301} {doi:
  10.1103/PhysRevA.84.050301}

\bibitem{James01}
James DFV, Kwiat PG, Munro WJ, White AG. Measurement of qubits. {\it Phys. Rev.
  A} 2001\string; 64\string: 052312.
\newblock \href {\doibase 10.1103/PhysRevA.64.052312} {doi:
  10.1103/PhysRevA.64.052312}

\bibitem{Lvovsky_2004}
Lvovsky AI. Iterative maximum-likelihood reconstruction in quantum homodyne
  tomography. {\it Journal of Optics B: Quantum and Semiclassical Optics}
  2004\string; 6(6)\string: S556--S559.
\newblock \href {\doibase 10.1088/1464-4266/6/6/014} {doi:
  10.1088/1464-4266/6/6/014}

\bibitem{Anis_2012}
Anis A, Lvovsky AI. Maximum-likelihood coherent-state quantum process
  tomography. {\it New Journal of Physics} 2012\string; 14(10)\string: 105021.
\newblock \href {\doibase 10.1088/1367-2630/14/10/105021} {doi:
  10.1088/1367-2630/14/10/105021}

\bibitem{spag17sr}
Spagnolo N, Maiorino E, Vitelli C, et al. Learning an unknown transformation
  via a genetic approach. {\it Scientific Reports} 2017\string; 7(1)\string:
  14316.
\newblock \href {\doibase 10.1038/s41598-017-14680-7} {doi:
  10.1038/s41598-017-14680-7}

\bibitem{gao18prl}
Gao J, Qiao LF, Jiao ZQ, et al. Experimental Machine Learning of Quantum
  States. {\it Phys. Rev. Lett.} 2018\string; 120\string: 240501.
\newblock \href {\doibase 10.1103/PhysRevLett.120.240501} {doi:
  10.1103/PhysRevLett.120.240501}

\bibitem{torlai18nat}
Torlai G, Mazzola G, Carrasquilla J, Troyer M, Melko R, Carleo G.
  Neural-network quantum state tomography. {\it Nature Physics} 2018\string;
  14(5)\string: 447--450.
\newblock \href {\doibase 10.1038/s41567-018-0048-5} {doi:
  10.1038/s41567-018-0048-5}

\bibitem{rocc19sa}
Rocchetto A, Aaronson S, Severini S, et al. Experimental learning of quantum
  states. {\it Science Advances} 2019\string; 5(3)\string: eaau1946.
\newblock \href {\doibase 10.1126/sciadv.aau1946} {doi: 10.1126/sciadv.aau1946}

\bibitem{torlai19prl}
Torlai G, Timar B, Nieuwenburg vEPL, et al. Integrating Neural Networks with a
  Quantum Simulator for State Reconstruction. {\it Phys. Rev. Lett.}
  2019\string; 123\string: 230504.
\newblock \href {\doibase 10.1103/PhysRevLett.123.230504} {doi:
  10.1103/PhysRevLett.123.230504}

\bibitem{gior20prl}
Giordani T, Suprano A, Polino E, et al. Machine Learning-Based Classification
  of Vector Vortex Beams. {\it Phys. Rev. Lett.} 2020\string; 124\string:
  160401.
\newblock \href {\doibase 10.1103/PhysRevLett.124.160401} {doi:
  10.1103/PhysRevLett.124.160401}

\bibitem{palmieri20npj}
Palmieri AM, Kovlakov E, Bianchi F, et al. Experimental neural network enhanced
  quantum tomography. {\it npj Quantum Information} 2020\string; 6(1)\string:
  20.
\newblock \href {\doibase 10.1038/s41534-020-0248-6} {doi:
  10.1038/s41534-020-0248-6}

\bibitem{tiun20opt}
Tiunov ES, (Vyborova) VVT, Ulanov AE, Lvovsky AI, Fedorov AK. Experimental
  quantum homodyne tomography via machine learning. {\it Optica} 2020\string;
  7(5)\string: 448--454.
\newblock \href {\doibase 10.1364/OPTICA.389482} {doi: 10.1364/OPTICA.389482}

\bibitem{agre19prx}
Agresti I, Viggianiello N, Flamini F, et al. Pattern Recognition Techniques for
  Boson Sampling Validation. {\it Phys. Rev. X} 2019\string; 9\string: 011013.
\newblock \href {\doibase 10.1103/PhysRevX.9.011013} {doi:
  10.1103/PhysRevX.9.011013}

\bibitem{flam19qst}
Flamini F, Spagnolo N, Sciarrino F. Visual assessment of multi-photon
  interference. {\it Quantum Science and Technology} 2019\string; 4(2)\string:
  024008.
\newblock \href {\doibase 10.1088/2058-9565/ab04fc} {doi:
  10.1088/2058-9565/ab04fc}

\bibitem{gebh20prr}
Gebhart V, Bohmann M. Neural-network approach for identifying nonclassicality
  from click-counting data. {\it Phys. Rev. Research} 2020\string; 2\string:
  023150.
\newblock \href {\doibase 10.1103/PhysRevResearch.2.023150} {doi:
  10.1103/PhysRevResearch.2.023150}

\bibitem{cimini20prl}
Cimini V, Barbieri M, Treps N, Walschaers M, Parigi V. Neural Networks for
  Detecting Multimode Wigner Negativity. {\it Phys. Rev. Lett.} 2020\string;
  125\string: 160504.
\newblock \href {\doibase 10.1103/PhysRevLett.125.160504} {doi:
  10.1103/PhysRevLett.125.160504}

\bibitem{hent10prl}
Hentschel A, Sanders BC. Machine Learning for Precise Quantum Measurement. {\it
  Phys. Rev. Lett.} 2010\string; 104\string: 063603.
\newblock \href {\doibase 10.1103/PhysRevLett.104.063603} {doi:
  10.1103/PhysRevLett.104.063603}

\bibitem{cimi19prl}
Cimini V, Gianani I, Spagnolo N, Leccese F, Sciarrino F, Barbieri M.
  Calibration of Quantum Sensors by Neural Networks. {\it Phys. Rev. Lett.}
  2019\string; 123\string: 230502.
\newblock \href {\doibase 10.1103/PhysRevLett.123.230502} {doi:
  10.1103/PhysRevLett.123.230502}

\bibitem{cimini21pra}
Cimini V, Polino E, Valeri M, et al. Calibration of Multiparameter Sensors via
  Machine Learning at the Single-Photon Level. {\it Phys. Rev. Applied}
  2021\string; 15\string: 044003.
\newblock \href {\doibase 10.1103/PhysRevApplied.15.044003} {doi:
  10.1103/PhysRevApplied.15.044003}

\bibitem{hent11prl}
Hentschel A, Sanders BC. Efficient Algorithm for Optimizing Adaptive Quantum
  Metrology Processes. {\it Phys. Rev. Lett.} 2011\string; 107\string: 233601.
\newblock \href {\doibase 10.1103/PhysRevLett.107.233601} {doi:
  10.1103/PhysRevLett.107.233601}

\bibitem{love13prl}
Lovett NB, Crosnier C, Perarnau-Llobet M, Sanders BC. Differential Evolution
  for Many-Particle Adaptive Quantum Metrology. {\it Phys. Rev. Lett.}
  2013\string; 110\string: 220501.
\newblock \href {\doibase 10.1103/PhysRevLett.110.220501} {doi:
  10.1103/PhysRevLett.110.220501}

\bibitem{bona16nat}
Bonato C, Blok MS, Dinani HT, et al. Optimized quantum sensing with a single
  electron spin using real-time adaptive measurements. {\it Nature
  Nanotechnology} 2016\string; 11(3)\string: 247--252.
\newblock \href {\doibase 10.1038/nnano.2015.261} {doi: 10.1038/nnano.2015.261}

\bibitem{pali17neuro}
Palittapongarnpim P, Wittek P, Zahedinejad E, Vedaie S, Sanders BC. Learning in
  quantum control: High-dimensional global optimization for noisy quantum
  dynamics. {\it Neurocomputing} 2017\string; 268\string: 116-126.
\newblock \href {\doibase https://doi.org/10.1016/j.neucom.2016.12.087} {doi:
  https://doi.org/10.1016/j.neucom.2016.12.087}

\bibitem{liu17pra}
Liu J, Yuan H. Control-enhanced multiparameter quantum estimation. {\it Phys.
  Rev. A} 2017\string; 96\string: 042114.
\newblock \href {\doibase 10.1103/PhysRevA.96.042114} {doi:
  10.1103/PhysRevA.96.042114}

\bibitem{paes17prl}
Paesani S, Gentile AA, Santagati R, et al. Experimental Bayesian Quantum Phase
  Estimation on a Silicon Photonic Chip. {\it Phys. Rev. Lett.} 2017\string;
  118\string: 100503.
\newblock \href {\doibase 10.1103/PhysRevLett.118.100503} {doi:
  10.1103/PhysRevLett.118.100503}

\bibitem{lumi18pra}
Lumino A, Polino E, Rab AS, et al. Experimental Phase Estimation Enhanced by
  Machine Learning. {\it Phys. Rev. Applied} 2018\string; 10\string: 044033.
\newblock \href {\doibase 10.1103/PhysRevApplied.10.044033} {doi:
  10.1103/PhysRevApplied.10.044033}

\bibitem{pali19pra}
Palittapongarnpim P, Sanders BC. Robustness of quantum-enhanced adaptive phase
  estimation. {\it Phys. Rev. A} 2019\string; 100\string: 012106.
\newblock \href {\doibase 10.1103/PhysRevA.100.012106} {doi:
  10.1103/PhysRevA.100.012106}

\bibitem{dina19prb}
Dinani HT, Berry DW, Gonzalez R, Maze JR, Bonato C. Bayesian estimation for
  quantum sensing in the absence of single-shot detection. {\it Phys. Rev. B}
  2019\string; 99\string: 125413.
\newblock \href {\doibase 10.1103/PhysRevB.99.125413} {doi:
  10.1103/PhysRevB.99.125413}

\bibitem{liu20mlst}
Liu G, Chen M, Liu YX, Layden D, Cappellaro P. Repetitive readout enhanced by
  machine learning. {\it Machine Learning: Science and Technology} 2020\string;
  1(1)\string: 015003.
\newblock \href {\doibase 10.1088/2632-2153/ab4e24} {doi:
  10.1088/2632-2153/ab4e24}

\bibitem{peng20pra}
Peng Y, Fan H. Feedback ansatz for adaptive-feedback quantum metrology training
  with machine learning. {\it Phys. Rev. A} 2020\string; 101\string: 022107.
\newblock \href {\doibase 10.1103/PhysRevA.101.022107} {doi:
  10.1103/PhysRevA.101.022107}

\bibitem{ramb20prr}
Rambhatla K, D'Aurelio SE, Valeri M, Polino E, Spagnolo N, Sciarrino F.
  Adaptive phase estimation through a genetic algorithm. {\it Phys. Rev.
  Research} 2020\string; 2\string: 033078.
\newblock \href {\doibase 10.1103/PhysRevResearch.2.033078} {doi:
  10.1103/PhysRevResearch.2.033078}

\bibitem{knot16njp}
Knott PA. A search algorithm for quantum state engineering and metrology. {\it
  New Journal of Physics} 2016\string; 18(7)\string: 073033.
\newblock \href {\doibase 10.1088/1367-2630/18/7/073033} {doi:
  10.1088/1367-2630/18/7/073033}

\bibitem{arra19qst}
Arrazola JM, Bromley TR, Izaac J, Myers CR, Br{\'{a}}dler K, Killoran N.
  Machine learning method for state preparation and gate synthesis on photonic
  quantum computers. {\it Quantum Science and Technology} 2019\string;
  4(2)\string: 024004.
\newblock \href {\doibase 10.1088/2058-9565/aaf59e} {doi:
  10.1088/2058-9565/aaf59e}

\bibitem{nich19qst}
Nichols R, Mineh L, Rubio J, Matthews JCF, Knott PA. Designing quantum
  experiments with a genetic algorithm. {\it Quantum Science and Technology}
  2019\string; 4(4)\string: 045012.
\newblock \href {\doibase 10.1088/2058-9565/ab4d89} {doi:
  10.1088/2058-9565/ab4d89}

\bibitem{meln18pnas}
Melnikov AA, Poulsen~Nautrup H, Krenn M, et al. Active learning machine learns
  to create new quantum experiments. {\it Proceedings of the National Academy
  of Sciences} 2018\string; 115(6)\string: 1221--1226.
\newblock \href {\doibase 10.1073/pnas.1714936115} {doi:
  10.1073/pnas.1714936115}

\bibitem{kren16prl}
Krenn M, Malik M, Fickler R, Lapkiewicz R, Zeilinger A. Automated Search for
  new Quantum Experiments. {\it Phys. Rev. Lett.} 2016\string; 116\string:
  090405.
\newblock \href {\doibase 10.1103/PhysRevLett.116.090405} {doi:
  10.1103/PhysRevLett.116.090405}

\bibitem{dris19qmi}
O'Driscoll L, Nichols R, Knott PA. A hybrid machine learning algorithm for
  designing quantum experiments. {\it Quantum Machine Intelligence}
  2019\string; 1(1)\string: 5--15.
\newblock \href {\doibase 10.1007/s42484-019-00003-8} {doi:
  10.1007/s42484-019-00003-8}

\bibitem{saba19pra}
Sabapathy KK, Qi H, Izaac J, Weedbrook C. Production of photonic universal
  quantum gates enhanced by machine learning. {\it Phys. Rev. A} 2019\string;
  100\string: 012326.
\newblock \href {\doibase 10.1103/PhysRevA.100.012326} {doi:
  10.1103/PhysRevA.100.012326}

\bibitem{gao20prl}
Gao X, Erhard M, Zeilinger A, Krenn M. Computer-Inspired Concept for
  High-Dimensional Multipartite Quantum Gates. {\it Phys. Rev. Lett.}
  2020\string; 125\string: 050501.
\newblock \href {\doibase 10.1103/PhysRevLett.125.050501} {doi:
  10.1103/PhysRevLett.125.050501}

\bibitem{jaeg21nce}
Jaeger H. Towards a generalized theory comprising digital, neuromorphic and
  unconventional computing. {\it Neuromorphic Computing and Engineering}
  2021\string; 1(1)\string: 012002.
\newblock \href {\doibase 10.1088/2634-4386/abf151} {doi:
  10.1088/2634-4386/abf151}

\bibitem{muja21}
Mujal P, Mart{\'\i}nez-Pe{\~n}a R, Nokkala J, et al. Opportunities in Quantum
  Reservoir Computing and Extreme Learning Machines. {\it Advanced Quantum
  Technologies} 2021\string; 4(8)\string: 2100027.
\newblock \href {\doibase https://doi.org/10.1002/qute.202100027} {doi:
  https://doi.org/10.1002/qute.202100027}

\bibitem{Schmale22}
Schmale T, Reh M, G\"arttner M. Efficient quantum state tomography with
  convolutional neural networks. {\it npj Quantum Information} 2022\string;
  8\string: 115.
\newblock \href {\doibase 10.1038/s41534-022-00621-4} {doi:
  10.1038/s41534-022-00621-4}

\bibitem{PhysRevA.63.042304}
Fujiwara A. Quantum channel identification problem. {\it Phys. Rev. A}
  2001\string; 63\string: 042304.
\newblock \href {\doibase 10.1103/PhysRevA.63.042304} {doi:
  10.1103/PhysRevA.63.042304}

\bibitem{PhysRevApplied.13.034013}
Lo HP, Ikuta T, Matsuda N, Honjo T, Munro WJ, Takesue H. Quantum Process
  Tomography of a Controlled-Phase Gate for Time-Bin Qubits. {\it Phys. Rev.
  Appl.} 2020\string; 13\string: 034013.
\newblock \href {\doibase 10.1103/PhysRevApplied.13.034013} {doi:
  10.1103/PhysRevApplied.13.034013}

\bibitem{White:07}
White AG, Gilchrist A, Pryde GJ, O'Brien JL, Bremner MJ, Langford NK. Measuring
  two-qubit gates. {\it J. Opt. Soc. Am. B} 2007\string; 24(2)\string:
  172--183.
\newblock \href {\doibase 10.1364/JOSAB.24.000172} {doi:
  10.1364/JOSAB.24.000172}

\bibitem{goodfellow16}
Goodfellow I, Bengio Y, Courville A. {\it Deep Learning}.
\newblock MIT Press .
\newblock 2016.
\newblock \url{http://www.deeplearningbook.org}.

\bibitem{Hashisho2019}
Hashisho Y, Albadawi M, Krause T, {Von Lukas} UF. Underwater color restoration
  using u-net denoising autoencoder.  2019\string; 2019-September\string:
  117--122.
\newblock \href {\doibase 10.1109/ISPA.2019.8868679} {doi:
  10.1109/ISPA.2019.8868679}

\bibitem{chiuri11prl}
Chiuri A, Rosati V, Vallone G, et al. Experimental Realization of Optimal Noise
  Estimation for a General Pauli Channel. {\it Phys. Rev. Lett.} 2011\string;
  107\string: 253602.
\newblock \href {\doibase 10.1103/PhysRevLett.107.253602} {doi:
  10.1103/PhysRevLett.107.253602}

\bibitem{Alteprl03}
Altepeter JB, Branning D, Jeffrey E, et al. Ancilla-Assisted Quantum Process
  Tomography. {\it Phys. Rev. Lett.} 2003\string; 90\string: 193601.
\newblock \href {\doibase 10.1103/PhysRevLett.90.193601} {doi:
  10.1103/PhysRevLett.90.193601}

\bibitem{mohseni08pra}
Mohseni M, Rezakhani AT, Lidar DA. Quantum-process tomography: Resource
  analysis of different strategies. {\it Phys. Rev. A} 2008\string; 77\string:
  032322.
\newblock \href {\doibase 10.1103/PhysRevA.77.032322} {doi:
  10.1103/PhysRevA.77.032322}

\end{thebibliography}

\end{document}